# Analysing the linearised radially polarised light source for improved precision in strain measurement using micro-Raman spectroscopy

V. Prabhakara[1,2,3], T. Nuytten[2], H. Bender[2], W. Vandervorst[2,4], S. Bals[1,3] and J. Verbeeck[1,3]

1. EMAT, University of Antwerp, Groenenborgerlaan 171, 2020, Antwerp, Belgium.

2. Imec, Kapeldreef 75, 3001, Leuven, Belgium.

3. NANOlab Center of Excellence, University of Antwerp, Belgium

4. Instituut voor Kern-en Stralingsfysica, KU Leuven, B-3001 Leuven, Belgium.

## Abstract:

Strain engineering in semiconductor transistor devices has become vital in the semiconductor industry due to the ever increasing need for performance enhancement at the nanoscale. Raman spectroscopy is a non-invasive measurement technique with high sensitivity to mechanical stress that does not require any special sample preparation procedures in comparison to characterization involving transmission electron microscopy (TEM), making it suitable for inline strain measurement in the semiconductor industry. Indeed at present, strain measurements using Raman spectroscopy are already routinely carried out in semiconductor devices as it is cost effective, fast and non-destructive. In this paper we explore the usage of linearised-radially polarised light as an excitation source, which does provide significantly enhanced accuracy and precision as compared to linearly polarised light for this application. Numerical simulations are done to quantitatively evaluate the electric field intensities that contribute to this enhanced sensitivity. We benchmark the experimental results against TEM diffraction-based techniques like nano-beam diffraction and Bessel diffraction. Differences between both approaches are assigned to strain relaxation due to sample thinning required in TEM setups, demonstrating the benefit of Raman for nondestructive inline testing.

## Introduction:

Scaling down the size of transistors has come at a cost due to issues like short channel effects, increased leakage current, decreased reliability and low yield [1,2].To overcome these limitations, FinFETs (Fin Field Effect Transistors) are introduced as 3D transistors mitigating the short channel effects in planar MOSFETs by providing enhanced electrostatic control over the channel, better switching characteristics with a reasonable increase in the production cost[3]. Strain engineering has proven to increase the carrier mobility, which in turn extends the scaling limits and improves the electrical performance[4]. Most commonly this is done by introducing Si-Ge into the channel and/or source/drain regions. In essence, strain modifies the band structure of silicon-germanium materials and their alloys in a way that compressive strain increases the hole mobility in p-MOS transistors and tensile strain increases the electron mobility in n-MOS transistors[5,6]. Hence, monitoring strain during the development and production process with good precision and accuracy is vital for the semiconductor industry.

Transmission electron microscopy techniques provide the best spatial resolution for the latest semiconductor technology nodes to measure strain at the nanoscale. There are a plethora of strain measurement techniques including HR-STEM and moiré[7–10], that provide images of the material at the nanoscale. Further processing these images with techniques like geometric phase analysis[11] and peak-pairs analysis[12] can extract the strain with very high spatial resolution of less than 1 nm and good precision of $1 \times 10^{-3}$. Diffraction-based techniques like nano-beam diffraction (NBD) with and without precession[13–17] provide good spatial resolution of 1 - 6 nm, a very high precision of $2 \times 10^{-4}$ - $6 \times 10^{-4}$ and an accuracy of $1 \times 10^{-3}$. Bessel diffraction[18] provides a spatial resolution between 1- 3 nm, excellent precision of $2.5 \times 10^{-4}$ and an accuracy of $1.5 \times 10^{-3}$. As such, these diffraction-based techniques can be used as a benchmark for strain analysis using TEM at the nanoscale. Nevertheless in all of these cases a concern is present that the required sample preparation leads to strain relaxation within the thin lamella and thus to an underestimation of the real strain[19].

Raman spectroscopy provides an alternative and indirect way to measure strain in Si-Ge semiconductor materials. The shift in the observed phonon peaks is associated with the stress present in the material and is directly dependent on the phonon deformation potentials[20,21]. The strain can then be calculated using Hooke's law of elasticity under the elastic regime of stress-strain behaviour, which is typical for an epitaxially strained Si-Ge nanodevice. Raman spectroscopy is usually performed in the backscattering configuration using a laser beam as the excitation source. Although the diffraction-limited spatial resolution in the visible wavelength range of the laser source is lower than the device dimension which needs to be probed, the technique remains applicable when employing the concept of nano-focussing[22,23]. In this concept, a parallel array of nanodevices is probed which actually form a waveguide-like structure that only allows specific transmission modes depending on the polarization conditions, leading to an enhancement of the local electric field and hence amplifying the Raman signal from the nanodevices [24]. The signal collected from such an array of nanodevices is the sum over many individual devices and the stress calculated is the average stress over the probed location. Hence, it does not provide details on the nanoscale spatial strain distribution inside one particular nanodevice. However, averaging over multiple devices is an asset as it improves the statistical significance of the results as long as the inter device variation is kept small.

In this paper, we measure stress and calculate strain in an epitaxially-strained Ge structure and a 16 nm-wide finFET structure using Raman spectroscopy. Initial experiments have shown that transverse optical (TO) and longitudinal optical (LO) phonon modes related to in-plane stress in $\sigma_{xx}$ and $\sigma_{yy}$ are very difficult to quantify separately due to their spectral overlap. In order to overcome this experimental difficulty we introduce a radially polarised setup that enables to increase the TO (transverse optical phonon) contribution relative to the LO contribution by increasing the longitudinal electric field component in the stimulus and we show that this setup can significantly enhance the precision of the obtained strain results.

Experimentally, we use two types of Raman excitation sources: Linearly polarised and linearised - radially polarised laser light. The radial polarisation in this experiment is achieved by using an S-waveplate (manufactured by wophotonics, Lithuania) which directly converts the incoming linearly polarised light to radially polarised light. Radially polarised light in general provides a stronger longitudinal component of the electric field in comparison to linearly polarised light[25,26]. This increase in the longitudinal component of the electric field is expected to increase the TO component in comparison to the LO component which enables a more sensitive determination of the parameters of the phonon peaks that partially overlap . We quantitatively evaluate the impact of the use of these excitation sources through electric field simulations and calculate the respective electric field distributions at the focal plane and the obtainable spatial resolution in the cartesian coordinate system and their impact on the Raman based stress measurements. We also show the effects of the excitation source on the obtained Raman spectrum from strained Ge and 16 nm  finFET arrays. Finally, the calculated strain values are compared against the strain values measured with the TEM NBD and Bessel diffraction techniques.

## Raman stress measurement:

Raman spectroscopy is a non-destructive tool to characterize a variety of material parameters like for instance mechanical stress in crystalline materials. It is used primarily to study the vibrational energy levels in a crystal using laser light as an excitation source.

The Raman scattering efficiency $I$ is proportional to the incident light polarization and the outgoing light polarization and is given by [27,28]

$$I = C\sum_{j} \left| e_{out}^{T} R_j e_{in} \right|^2 \qquad (1)$$

where, C is a constant and $R_j$ is the Raman tensor for the $j^{th}$ phonon, $e_{out}$ and $e_{in}$ are the outgoing and incoming polarisation vectors. The subscript T corresponds to the transposed vector. The Raman tensors are obtained from group theoretical considerations which are used to calculate the polarization selection rules[29]. In zinc blende-type or diamond-type point group semiconductors (silicon, germanium), the Raman tensors in the cartesian coordinate system x = [100], y = [010] and z = [001] are given by

$$R_1 = \begin{bmatrix} 0 & 0 & 0 \\ 0 & 0 & d \\ 0 & d & 0 \end{bmatrix}, R_2 = \begin{bmatrix} 0 & 0 & d \\ 0 & 0 & 0 \\ d & 0 & 0 \end{bmatrix}, R_3 = \begin{bmatrix} 0 & d & 0 \\ d & 0 & 0 \\ 0 & 0 & 0 \end{bmatrix} \quad (2)$$

Hence, there is one longitudinal optical phonon (LO) and two transverse optical phonons (TO) in the Si and Ge crystals resulting in a total of three active Raman peaks at k = 0 (central point of the Brillouin zone). The LO and TO phonons are identified by the polarisation states of the incoming excitation light and the outgoing Raman scattered light, which based on eq. 1 gives rise to particular phonons. For example, considering the incoming and outgoing/Raman scattered light polarization both to be in the x direction ($e_{in} = [1\ 0\ 0]^T$) i.e., referring to the LO phonon excitation, solving eq.1 shows that $R_2$ links to LO phonon scattering. Similarly, $R_{1,3}$ links to TO phonons scattering. The three phonons are degenerate in the bulk crystal due to the crystal symmetry but the introduction of strain alters the symmetry and lifts this degeneracy[30]. Stress measurements using Raman spectroscopy are performed by measuring the shift of the center of these phonon peaks in comparison to their location in the unstrained material. The relation between the shift in the phonon peaks $\Delta\omega_{LO-TO}$ relative to the position in the degenerate bulk and the normal stresses $\sigma_{x,y}$ is given by [31]:

$$\Delta\omega_{LO} = \frac{1}{2\omega_0}\left(pS_{12} + q(S_{11} + S_{12})\right)\left(\sigma_x + \sigma_y\right) \quad (3)$$

$$\Delta\omega_{TO_1} = \frac{1}{2\omega_0}[(\frac{1}{2}(S_{11}+S_{12})(p+q) + qS_{12} + \frac{r}{2}S_{44})\sigma_x \\ + (\frac{1}{2}(S_{11}+S_{12})(p+q) + qS_{12} - \frac{r}{2}S_{44})\sigma_y] \quad (4)$$

$$\Delta\omega_{TO_2} = \frac{1}{2\omega_0}[(\frac{1}{2}(S_{11}+S_{12})(p+q) + qS_{12} - \frac{r}{2}S_{44})\sigma_x \\ + (\frac{1}{2}(S_{11}+S_{12})(p+q) + qS_{12} + \frac{r}{2}S_{44})\sigma_y] \quad (5)$$

Where, $S_{ij}$ are the components of the compliance matrix, p,q and r are the phonon deformation potentials, and $\omega_0$ is the stress-free value for the Raman shift. The measured stress values are used to calculate the strain using Hooke's law of elasticity[32].

The semiconductor devices in our study are oriented in the $x_s$ = [110], $y_s$ = [− 110] and $z_s$ = [001] crystal coordinate system and it is of interest to consider this crystal coordinate system instead of the Cartesian, as it then becomes straightforward to interpret the calculated stress values directly in the crystal coordinate system. From eq.1 and 2, we can derive the polarization requirements for the incoming and the outgoing light in exciting particular phonon modes (Table 1). Since the Raman scattering experiment is done in backscattering configuration, we consider $z_s$ = [001] as the light propagation direction. $x_s$, $y_s$, $z_s$ are represented as x, y and z for simplicity. The experimental configuration is denoted in the Porto notation (excitation axis(excitation polarization analyser orientation)Raman scattering axis). For example, the notation $z(xy)\bar{z}$ denotes that $z$ and $\bar{z}$ are the directions of propagation of the exciting (incoming) and Raman scattered (outgoing) light, $x$ and $y$ are the polarization directions of the exciting and Raman scattered light respectively.

*Table 1 Raman phonon selectivity for different experimental configurations considering x = [110], y = [− 110] and z = [001] .*

| Experimental configuration (porto notation) | LO | $TO_1$ | $TO_2$ |
|---|---|---|---|
| $z(xx)\bar{z}$ | $d^2$ | 0 | 0 |
| $z(yy)\bar{z}$ | $d^2$ | 0 | 0 |
| $z(zx)\bar{z}$ | 0 | 0 | $d^2$ |
| $z(zy)\bar{z}$ | 0 | $d^2$ | 0 |
| $z(zz)\bar{z}$ | 0 | 0 | 0 |
| $z(xy)\bar{z}$ | 0 | 0 | 0 |
| $z(yx)\bar{z}$ | 0 | 0 | 0 |

The light polarization is transverse and restricted to the x and the y axis and hence in the backscattering configuration, one can excite only LO phonons. The introduction of high numerical aperture (NA) objective lenses and oil immersion lenses provides a high convergence angle and increases the z component of the electric field near the focal plane and thus enables the excitation of the TO phonons that are traditionally forbidden in the backscattering configuration[26]. Now, considering the backscattering experimental mode with the oil immersion lens (high numerical aperture(NA) of 1.4) and a linearly x-polarised light (polarised in x direction), the incoming light will have primarily the x- and some y- and z-polarised components due to the depolarisation effect[33,34] occurring in these types of optics. So, to excite the TO phonons (TO mode), the analyser is set to the y-direction (perpendicular to the incoming light polarisation) and with a high NA aperture, we encounter $z(xy)\bar{z}$, $z(zy)\bar{z}$ and $z(yy)\bar{z}$ scenarios. The $z(xy)\bar{z}$ configuration is Raman inactive, while $z(zy)\bar{z}$ excites the $TO_1$ phonon. The small y component present in highly convergent incident laser light results in $z(yy)\bar{z}$ and in combination with non-perfect analyzer optics (which should block this component in the current configuration) gives rise to a noticeable excitation of the LO phonon. Thus in the TO mode, we excite both the $TO_1$ and the LO phonons. In the LO mode configuration, the analyser is set parallel to the x-direction i.e. parallel to the polarisation of the incoming light source. In this scenario, we encounter $z(xx)\bar{z}$, $z(yx)\bar{z}$ and $z(zx)\bar{z}$ given the high NA oil immersion lens. The $z(yx)\bar{z}$ configuration is Raman inactive. The $TO_2$ which is excited by $z(zx)\bar{z}$ is concealed by the LO component from $z(xx)\bar{z}$ due to a highly intense x-component of incoming light (further details are provided in the electric field simulation section later). Hence, with backscattering Raman spectroscopy, we can excite one LO phonon and one TO phonon facilitating the measurement of biaxial in-plane stresses in the Si and Ge semiconductor materials as will be discussed below.

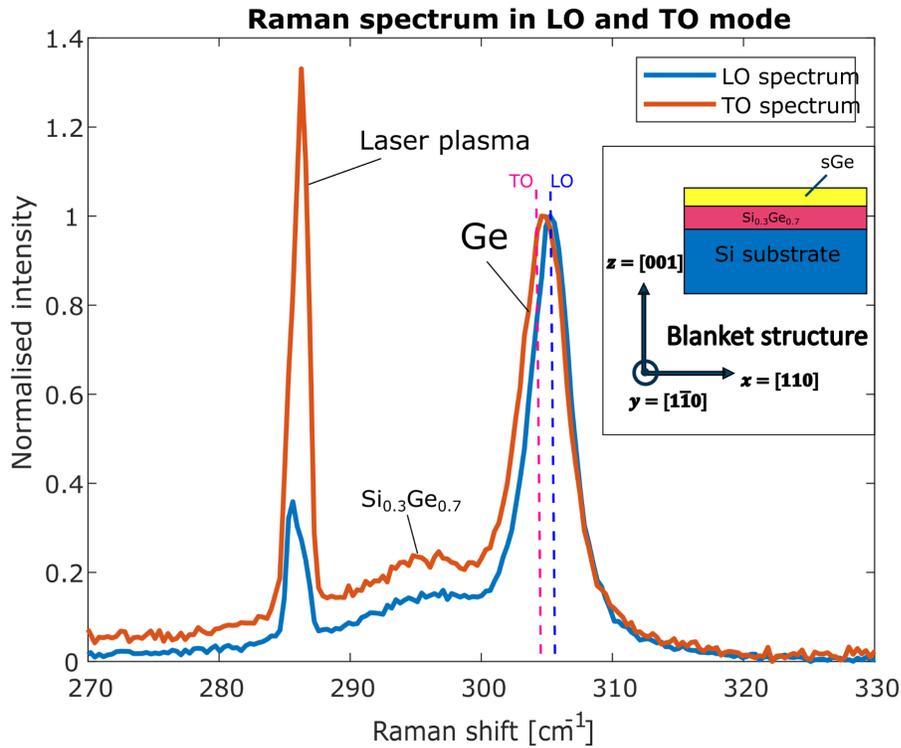

*Figure 1 Experimental Raman spectra on SiGe blanket structure (see text for detailed explanation on the material) consisting of the strained Ge region. The spectra are normalized with respect to the Ge peak. Note the broad Ge peak in the TO mode clearly indicates the presence of both TO and LO peaks. The dotted lines indicate the expected theoretical peak positions of TO and LO in the strained Ge showing the difficulty to resolve them as separate peaks in the experimental setup.*

Figure 1 illustrates the experimental Raman spectra taken on a blanket structure consisting of a thick $Si_{0.3}Ge_{0.7}$ layer grown on a Si substrate (Figure 1 inset). The $Si_{0.3}Ge_{0.7}$ layer is completely relaxed and acts as a strain relaxed buffer (SRB). A thin layer of Ge is epitaxially grown on top of the SRB and is expected to be compressively biaxially strained due to the lattice mismatch between $Si_{0.3}Ge_{0.7}$ and Ge. Notice that the TO and LO components of the Ge peak in the TO mode overlap each other strongly leading to the broad Ge peak. Though one can use the prior knowledge from the LO mode spectrum (that predominantly consists of the LO phonon) to estimate the peak shape of the LO-contribution, separating LO/TO unambiguously remains very difficult. The required process to disentangle these peaks would become more reliable if one could vary the intensity of the TO component more(relative to the LO component) in the overlapping spectrum.

To achieve this, we explore the use of radially polarised light. The latter has a strong and centrally focused longitudinal component of light compared to linearly polarised light. This boosts the intensity of the TO peaks in the Raman spectra in the TO mode, relative to the LO peak, which would allow us to separate LO and TO in a more reliable way thus providing insight into the detailed biaxial stress distributions. To illustrate the efficiency of our approach we study blanket structures and 16 nm finFETs.

In the next section, we will compare the effect of using the different Raman excitation sources (i.e. linearly, radially and linearised - radially polarised light) through numerical simulations with the purpose to investigate their potential effect on enhanced sensitivity for the detection of TO phonons in the TO mode.

# Numerical simulation of the electric field components at the focal plane

### a) Linearly and radially polarised light

The study on the electric field intensity distribution of the radially polarised light has been a topic of interest in the last few decades. Radially polarised light has a strong and centrally focussed longitudinal component of light compared to linearly polarised light. This boosts the intensity of the TO peaks in the Raman spectra in the TO mode, which are usually accompanied by the LO peak. The intensity distribution for the radially polarised light in the focal plane is given by[35]:

$$Ix(\rho, z) = -\frac{iA}{\pi} \int_0^{2\pi} \int_0^{\alpha} \cos^{1/2}\theta \, \sin\theta \, \cos\theta \, \cos(\varphi) I_o(\theta) e^{ik(z\cos\theta + \rho\sin\theta\cos(\varphi - \varphi_s))} d\varphi d\theta \qquad (6)$$

$$Iy(\rho, z) = -\frac{iA}{\pi} \int_0^{2\pi} \int_0^\alpha \cos^{1/2}\theta \sin\theta \cos\theta \sin(\varphi) I_o(\theta) e^{ik(z\cos\theta + \rho\sin\theta\cos(\varphi-\varphi_s))} d\varphi d\theta \quad (7)$$

$$Ir(\rho, z) = -\frac{iA}{\pi} \int_0^{2\pi} \int_0^\alpha \cos^{1/2}\theta \sin\theta \cos\theta \cos(\varphi - \varphi_s) I_o(\theta) e^{ik(z\cos\theta + \rho\sin\theta\cos(\varphi-\varphi_s))} d\varphi d\theta \quad (8)$$

$$Iz(\rho, z) = -\frac{iA}{\pi} \int_0^{2\pi} \int_0^\alpha \cos^{1/2}\theta \sin^2\theta I_o(\theta) e^{ik(z\cos\theta + \rho\sin\theta\cos(\varphi-\varphi_s))} d\varphi d\theta \quad (9)$$

where, $A$ is a constant, $Ix, y$ are the electric field transverse components, $Ir$ and $Iz$ are the radial (transverse) and longitudinal components, $\rho$, $\varphi_s$ and $z$ are the radial, azimuthal and vertical cylindrical coordinates with respect to the origin in the focal plane; x,y, and z are the cartesian coordinates in the focal plane, $\theta$ is the angle a sub-ray makes with the optical axis and $\theta$ is integrated from 0 to the convergence angle $\alpha$(acceptance angle), $\varphi$ is the azimuth angle subtended by the sub-ray onto the focal plane and is integrated from 0 to $2\pi$ radian. $I_o(\theta)$ is the Bessel-Gauss apodization function as described in [36,37].

$$I_0(\theta) = exp\left[-\beta^2\left(\frac{\sin\theta}{\sin\alpha}\right)^2 J_1\left(2\beta\frac{\sin\theta}{\sin\alpha}\right)\right] \quad (10)$$

Where $\beta$ is the ratio of pupil radius and beam waist. Likewise, the electric field distribution for the linearly polarized light in the cartesian coordinate system has been extensively discussed in [38].

To gain insight into the electric field distributions for the specific case of our experiments, we performed numerical simulations using the parameters of our experimental setup i.e.the laser excitation wavelength of 633 nm, a convergence angle (equal to acceptance angle for the backscattered configuration) of 70°, a beam waist of 0.68 mm, a working distance of 0.13 mm, a numerical aperture of 1.4 (using an oil objective), a focal length of 0.129 mm and a pupil radius of 0.18 mm. The results of these calculations are summarized in Figures 2 and 3. Compared to the electric field distribution of the linearly polarised light, we can conclude that the radially polarised light has a strong centrally focussed longitudinal component of light and the transverse x and y components form the two side lobes (Figure 2). Vice versa, the linearly polarised light has a strong centrally focussed component in the direction of polarisation x and the y, z components are distributed away from the center (Figure 3). The units for the electric field are in Volt/m, but the absolute values depend on the laser power. Hence the normalised intensity is shown for the purpose of electric field distribution visualisation. Similarly, the relative intensities of the electric field components are tabulated in Table 2. The intensity in general is calculated as the sum of the amplitude square of the electric field. The relative intensity of a component is calculated as the intensity of that particular component divided by the total intensity. It can clearly be seen from the table that for radially polarized light, the Ez component is much larger in comparison to linearly polarized light. Since it is this component that is responsible for the excitation of the TO phonon in TO mode acquisition [$z(zy)\bar{z}$], our calculations show that for our experimental setup, we can expect a higher TO phonon intensity in the TO mode.

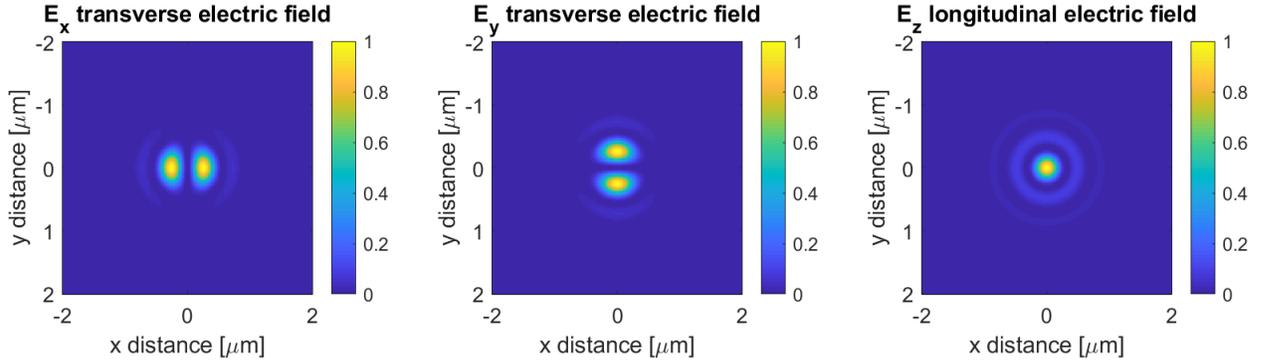

*Figure 2 Normalised electric field distribution of radially polarised light at the focal plane.*

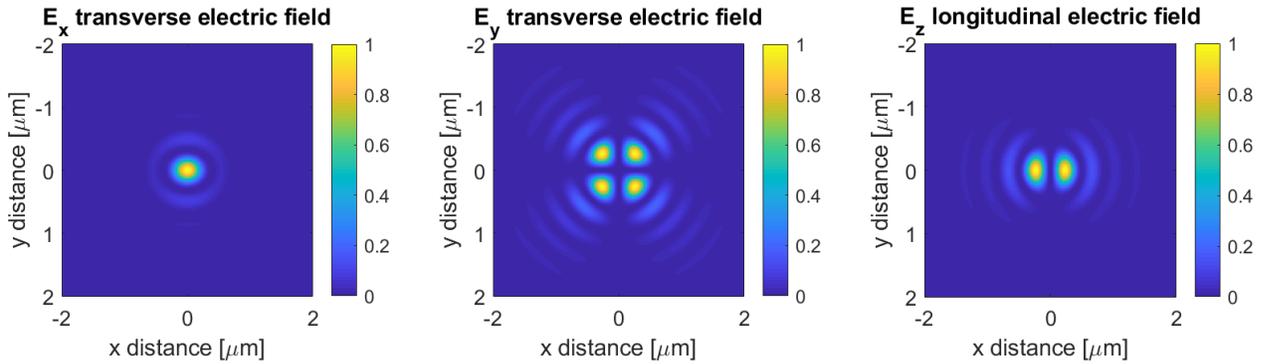

*Figure 3 Normalised electric field distribution of linearly polarised light at the focal plane.*

*Table 2 Relative intensity contributions of electric field components for linearly and radially polarised light at the focal plane.*

| Polarization | $E_x$ | $E_y$ | $E_z$ |
|---|---|---|---|
| Linear in x | 0.67 | 0.02 | 0.31 |
| Radial | 0.195 | 0.195 | 0.61 |

Finally, the normalised intensity distribution of the Ex, Ey and Ez components is calculated as a function of the radial distance from the centre, by radial integration and normalization to the total intensity in the distribution. This gives the normalised intensity distribution as a function of the diametric distance (Figure 4). The spatial resolution is calculated as the diameter of the disc that contains 68.5% of the total intensity (or one standard deviation from the center). The results are summarized in Table 3 and from here it can be seen that the radially polarised light provides a higher spatial resolution for both the longitudinal and transverse components and hence provides a tighter focus of the laser beam in the focal plane.

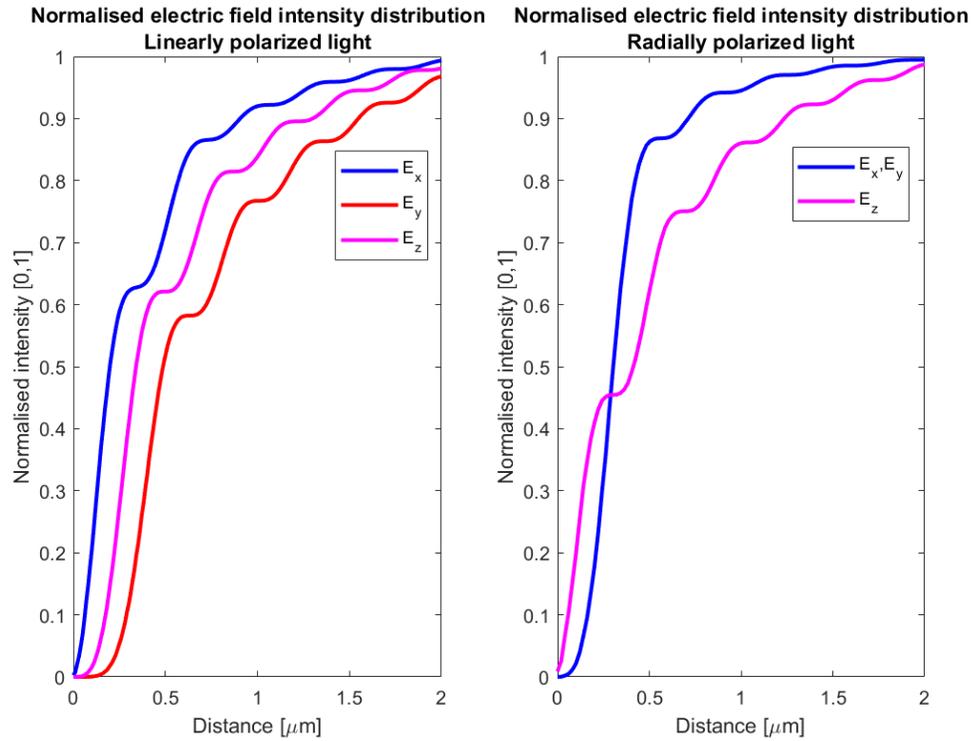

*Figure 4 Normalised intensity distribution of electric field components on the focal plane. The spatial resolution is the diametric distance or the diameter of the disc within which 68.5% of the total intensity lies.*

*Table 3 Comparison of obtainable spatial resolution for linearly and radially polarised light. The spatial resolution is calculated as the diametric distance or the diameter of the disc which contains 68.5% of the total intensity*

| Polarization | Spatial resolution [nm] | | |
|---|---|---|---|
| | $E_x$ | $E_y$ | $E_z$ |
| Linear in x | 455 | 818 | 728 |
| Radial | 364 | 364 | 545 |

We also calculate the relative intensities of the components as a function of the defocus distance (Figure 5). We notice a local enhancement of the longitudinal component of light close to the focal point (defocus = 0 µm) as expected for a high NA objective lens. The enhancement of the longitudinal component is very high in the case of radially polarised light in comparison to the linearly polarised light in line with the previous calculations. The Ex and Ey profiles remain identical for the radial setup.

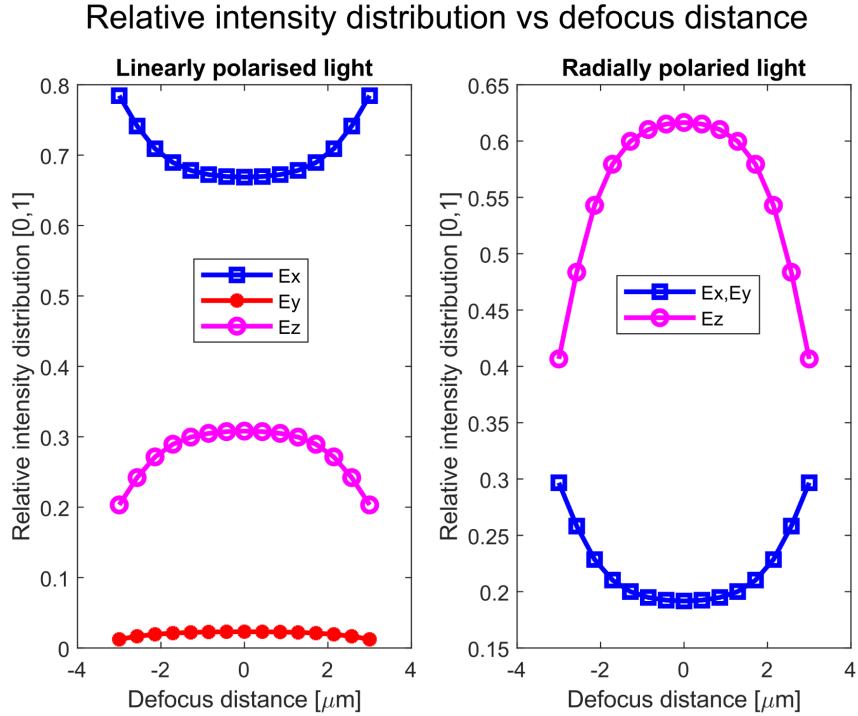

*Figure 5 Relative intensity distribution of electric field components as a function of defocus distance [μm] (defocus zero being the exact focal distance) for linearly (in x) and radially polarised light.*

## b) Linearised radially polarised light

In spite of the reduction in the transverse components for the radially polarised light, experimentally the TO phonon intensity appeared a lot weaker than the LO phonon in the TO mode (Figure 10b). This would result in a higher ambiguity in the identification of the TO peak resulting in highly unreliable stress measurements. For the $z(zy)\bar{z}$ configuration that results in the $TO_1$ excitation (Table 1), the Ey component of the incoming radially polarised light needs to be suppressed further in order to reduce the undesired excitation of LO phonon in the TO mode. In that regard, we have used a linearised - radially polarised setup [39] by placing a linear polariser after the radial polariser to further suppress the intensity of the Ey component of light (perpendicular to the linear polariser transmission axis) in the conventional radially polarised light at the focal plane. This helps to filter out or reduce the undesired component of light that is leading to LO phonon excitation while targeting the TO mode. We also numerically calculate the relative intensity distribution of the electric field components to evaluate the degree of reduction in the Ey (with x being the linear-polariser transmission axis) and also the effect on the spatial resolution from this modified setup.

The transmittance $T$ of a linear polariser is given as[40]:

$$T(\varphi) = (T_1 - T_2)cos^2\varphi + T_2 \qquad (11)$$

where, $\varphi$ is the angle between the polarisation of light and the transmission axis of the linear polariser, $T_1$ is the maximum transmittance value when the incoming light is parallel to the linear polariser transmission axis and $T_2$ is the minimum transmittance when the incoming light is perpendicular to the linear polariser transmission axis. Using eq.11 in eqs. 6 to 9 results in the electric field distribution equations for the linearised - radially polarised light:

$$Ix(\rho,z) = -\frac{iA}{\pi}\int_0^{2\pi}\int_0^\alpha T(\varphi)\cos^{1/2}\theta\,\sin\theta\,\cos\theta\,\cos(\varphi)I_o(\theta)e^{ik(z\cos\theta+\rho\sin\theta\cos(\varphi-\varphi_s))}d\varphi d\theta \quad (12)$$

$$Iy(\rho,z) = -\frac{iA}{\pi}\int_0^{2\pi}\int_0^\alpha T(\varphi)\cos^{1/2}\theta\,\sin\theta\,\cos\theta\,\sin(\varphi)I_o(\theta)e^{ik(z\cos\theta+\rho\sin\theta\cos(\varphi-\varphi_s))}d\varphi d\theta \quad (13)$$

$$Ir(\rho,z) = -\frac{iA}{\pi}\int_0^{2\pi}\int_0^\alpha T(\varphi)\cos^{1/2}\theta\,\sin\theta\,\cos\theta\,\cos(\varphi-\varphi_s)I_o(\theta)e^{ik(z\cos\theta+\rho\sin\theta\cos(\varphi-\varphi_s))} \quad (14)$$

$$Iz(\rho,z) = -\frac{iA}{\pi}\int_0^{2\pi}\int_0^\alpha T(\varphi)\cos^{1/2}\theta\,\sin^2\theta I_o(\theta)\,e^{ik(z\cos\theta+\rho\sin\theta\cos(\varphi-\varphi_s))}d\varphi d\theta \quad (15)$$

Figure 6 shows the normalised electric field distribution of the radial component $E_r$ at the focal plane for both the radially polarised light and the linearised - radially polarised light (with transmission axis of the linear polariser parallel to x - axis). The simulation parameters for the linear polariser include $T_1 = 0.83$ and extinction ratio $\rho_P = \frac{T_1}{T_2} = 800$ (in accordance with the specification for visible wire-grid polarizers). The radial component ($E_r$) of the radially polarised light is radially symmetric and possesses a doughnut-shaped distribution. Linearised - radially polarised light on the other hand has a skewed distribution where a higher intensity is concentrated in the direction of the transmission axis of the linear polariser (x-axis).

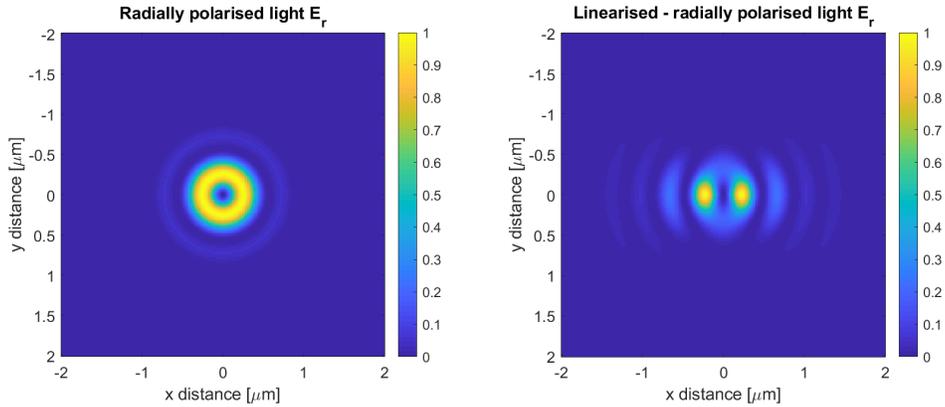

*Figure 6 Normalised intensity distribution of the radial electric field $E_r$ for a) radially polarised light and b) linearised - radially polarised light. The linear polariser transmission axis is parallel to the x axis.*

The electric field distributions of the x,y,z cartesian coordinates are shown in Figure 7. Here again, we observe a centrally focussed longitudinal component in comparison to the transverse components. The transverse components form the side lobes with field intensities concentrated away from the center.

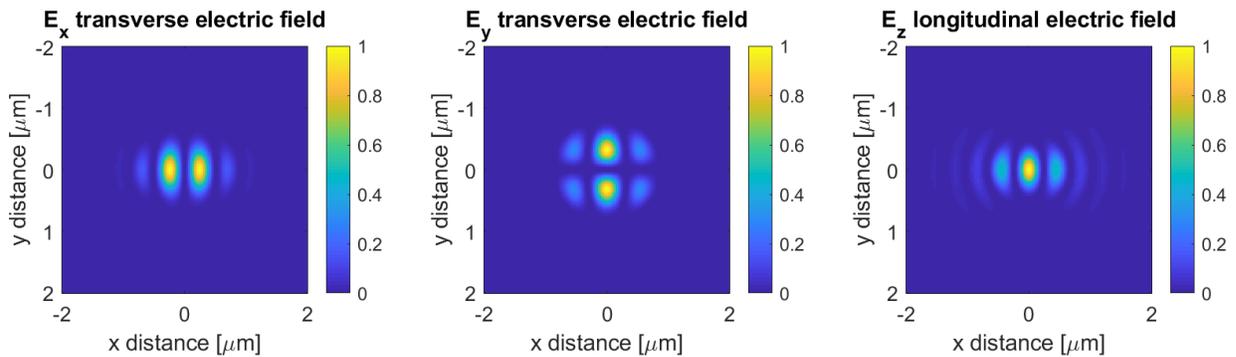

*Figure 7 Normalised electric field distribution of the linearised radially polarised light at the focal plane.*

The interest lies in the relative intensity contributions of the three electric field components (Table 4) Ex, Ey and Ez. The Ey component is greatly suppressed in comparison to the Ey component for radially polarised light (refer Table 2). This is because of the introduction of the linear polariser whose transmission axis is parallel to the x - axis. The longitudinal component is still higher and comparable to the radially polarised light setup. The relative intensity in Ex is higher in linearised - radially polarised light due to a higher transmission in the x-direction.

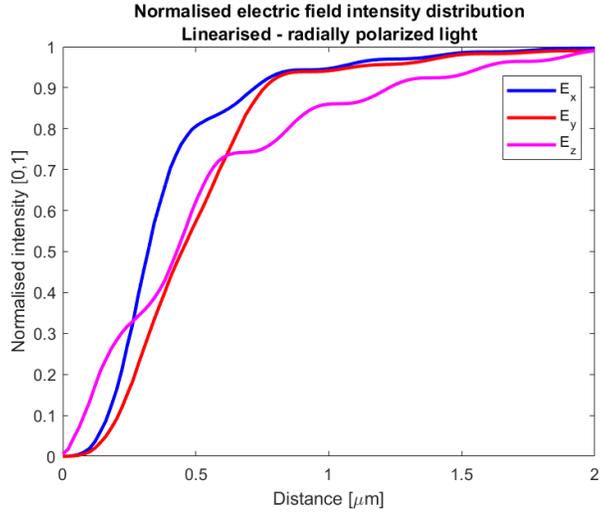

*Figure 8 Normalised intensity distribution of electric field components of the linearised-radially polarised light. The spatial resolution is the diametric distance within which 68.5% of the total intensity lies.*

*Table 4 Relative intensity and obtainable spatial resolution for linearised - radially polarised light. The spatial resolution is calculated as the diametric distance which contains 68.5% of the total intensity.*

| Parameter | $E_x$ | $E_y$ | $E_z$ |
|---|---|---|---|
| Relative intensity | 0.33 | 0.06 | 0.61 |
| Spatial resolution [nm] | 394 | 576 | 545 |

The benefit of the linearised - radially polarised light over the radially polarised light is the reduction in the Ey which is mainly responsible for the increased LO contribution in the TO mode ($z(zy)\,\bar{z}$). The Ex contribution, although higher than in the radially polarised light, is still half of the Ex contribution in the linearly polarised setup. This should lead to a reduction effect in the TO mode due to the transmission axis of the analyser placed in the y - direction which has a very high extinction coefficient ($\rho_P > 800$). Thus, a higher TO/LO ratio is expected for the linearised - radially polarised light for our experimental setup. The spatial resolution of the linearised - radially polarised light is comparable to the radially polarised setup in the Ex and Ez and a higher spatial spread of the Ey component is seen with the introduction of the linear polariser whose transmission axis is set parallel to x-axis.

The relative intensity as a function of the defocus distance (Figure 9) for the linearised - radially polarised light follows a similar trend as the radially polarised light for Ex and Ez, with a local enhancement of the longitudinal component Ez close to the focal plane (defocus = 0). Ex is relatively higher in the linearised - radially polarised due to the transmission axis of the linear polariser being aligned parallel to the x - direction and consequently Ey is greatly suppressed in comparison to the radially polarised light.

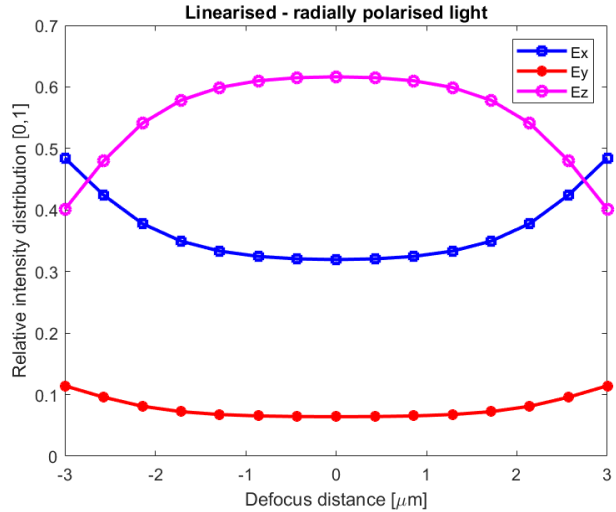

*Figure 9 Relative intensity distribution of electric field components as a function of the defocus distance [µm] (defocus zero being the exact focal distance) for linearised-radially polarised light.*

# Experimental results and discussion:

The Raman spectrum obtained in the TO configuration on a blanket structure is shown in Figure 10. It is clear that $z(zy)\bar{z}$ favours the $TO_1$ mode excitation (Table 1) and we show the spectra obtained using linearly polarised, radially polarised and linearised- radially polarised light (Figure 10a, 10b and 10c) . Due to the depolarization effect and the non-perfect optics, the LO peak will still be detectable and needs to be separated from the TO peak. For the $z(zy)\bar{z}$ configuration that results in the $TO_1$ excitation (Table 1), notice the strong LO component observed experimentally for the radially polarised light (Figure 10b) resulting in a higher ambiguity in the identification of the TO peak and further leading to highly unreliable stress measurements. A strong suppression of the LO is brought after linearisation due to reduction in the Ey component as observed for the linearised-radially polarised light (Figure 10c). Voigt profiles are used to fit the TO and LO peaks in Ge. We can clearly observe the difference in the expected peak positions for bulk, unstrained Ge ( dashed line in Figure 10a,b and c) versus the position for the strained Ge (sGe). The $Si_{0.3}Ge_{0.7}$ peak is fitted using an asymmetric function [41] which was found to give a better fit in comparison to a single Gaussian or a set of Gaussian and Lorentzian functions. Note that although the peak parameters for this peak are not used in our strain analysis it does lead to an optimal fitting for the other modes.

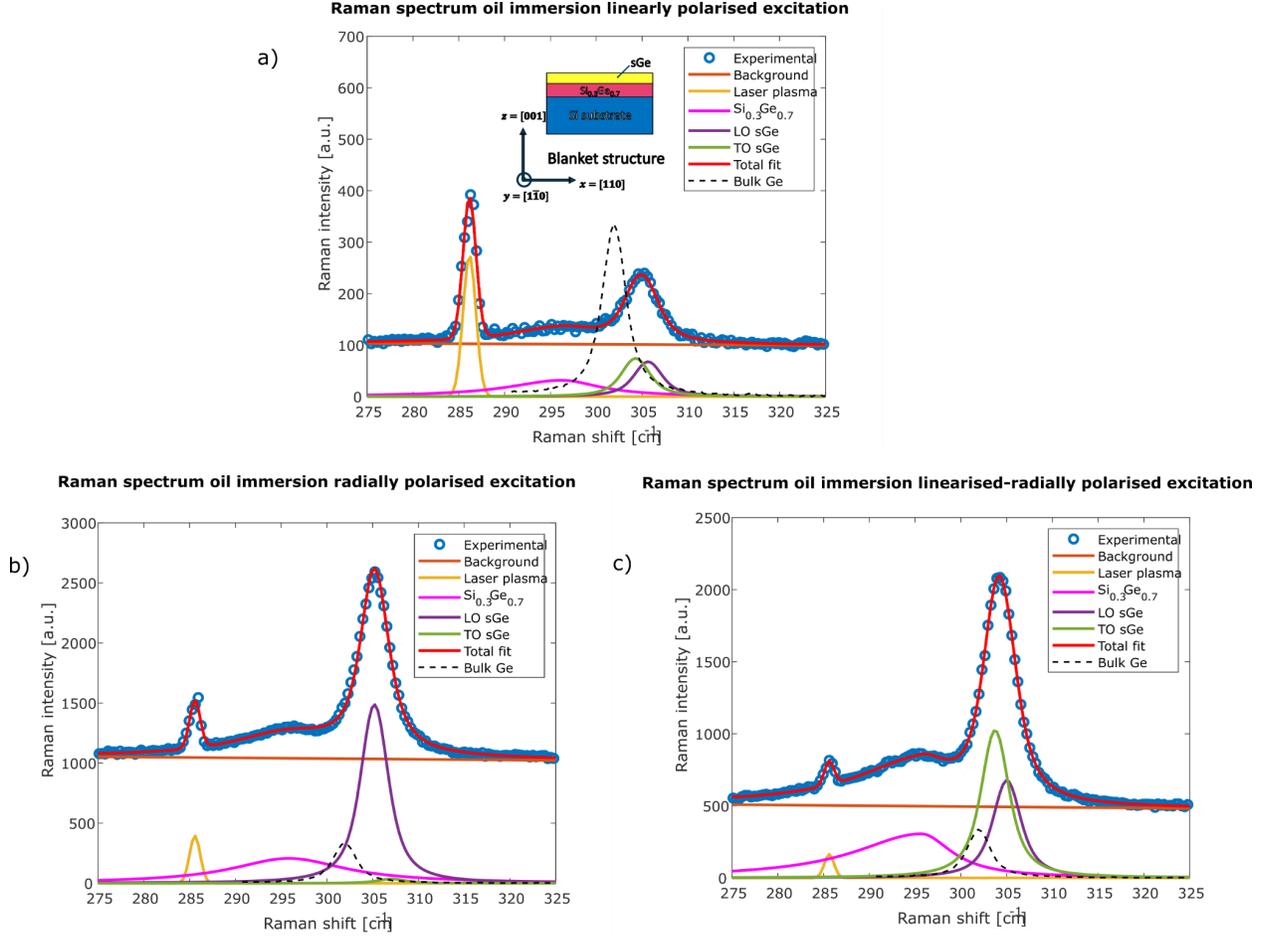

Figure 10 Raman spectrum of the Si-Ge blanket structure in the TO mode a) linearly polarised light , radially polarised light and c) linearised- radially polarised light. Notice the strong LO component observed experimentally for the radially polarised light and the strong suppression brought after linearisation as observed from the linearised-radially polarised light. Note the shift in the peak position with respect to the expected position for unstrained bulk Ge and the higher peak intensity of the sGe TO peak for the linearised-radially polarised light.

The asymmetric function $I$ as a function of the peak position $\omega$ in [cm$^{-1}$] is given by

$$I(\omega) = \frac{1}{2}\frac{[1 - sign(\omega - \omega_0)I_0]}{\left(\frac{\omega - \omega_0}{W_1}\right)^2 + 1} + \frac{1}{2}\frac{[1 - sign(\omega - \omega_0)I_0]}{\left(\frac{\omega_0 - \omega}{W_2}\right)^2 + 1} \qquad (9)$$

Where, $I_0$ is the peak intensity, $W_1$ and $W_2$ are the HWHM at the low frequency and high frequency side, $\omega_0$ is the peak position and $sign$ is the signum function of a real number. The Rayleigh scattered plasma peaks from the He-Ne laser are used as reference peaks to account for any instrumental drifts in the overall peak positions and these are fitted using Gaussian profiles.

The experimental procedure for measuring the stress is as follows: First the LO mode spectrum is acquired in the $z(xx)\bar{z}$ configuration, and the parameters of the LO peak (as fitted with a Voigt profile) of sGe are obtained. Similarly, the peak parameters of the asymmetric function fit for Si$_{0.3}$Ge$_{0.7}$ are obtained. These parameters are fixed while fitting the TO spectrum that is recorded in TO mode (implying use of a different combination of polarizers and analyzers in the experiment). This reduces the number of free parameters for the non-linear fitting or optimization algorithm and increases the accuracy of the fitting routine. The same parameters for the $Si_{0.3}Ge_{0.7}$ SRB are used in both the TO and LO spectrum due to the fact that strain is relaxed in the SRB meaning that the TO and LO peaks are degenerate [see Equations (3) - (5)]. We also fix the TO peak width to be the same as LO peak width. The peak/ line widths of the phonons are affected by the defects and/or impurities (doping) resulting in the asymmetric behaviour (as observed for the $Si_{0.3}Ge_{0.7}$)[42]. However, the stress introduced into the near defect-free semiconductor Ge channel would result in negligible variations to the line widths while predominantly influencing the shift in the peak position, justifying the usage of the same peak/line widths for TO and LO peaks in Ge[43]. The peak position is a free parameter necessary for stress calculation and is the variable in the optimization. Finally, equations (3) and (4) are used to solve for the two in-plane stresses $\sigma_x$ and $\sigma_y$. The Ge TO peaks show a higher peak intensity for the linearised-radially polarised setup (Figure 10c) in comparison to the linearly polarised setup due to the increase in the relative intensity of the longitudinal Ez component as summarised in the numerical simulation section. The TO/LO ratios (the TO/LO ratio has been calculated by individually measuring the areas of the TO and LO peaks) as calculated from multiple measurements for the linearly polarised light setup was 1.1 ± 0.2 and 1.6 ± 0.3 for the linearised-radially polarised setup. The error values quoted are the standard deviations of the ratios and can vary depending on the SNR in the data and hence an initial calibration of the laser intensity was made on a bulk Ge sample so that a good SNR is obtained while accounting for negligible thermal shifts in the phonon peaks. The data quality (or SNR) is assessed by calculating the uncertainty in the estimated peak position from Levenberg-Marquardt optimization for the bulk Ge Raman spectrum fitted with Voigt profile.

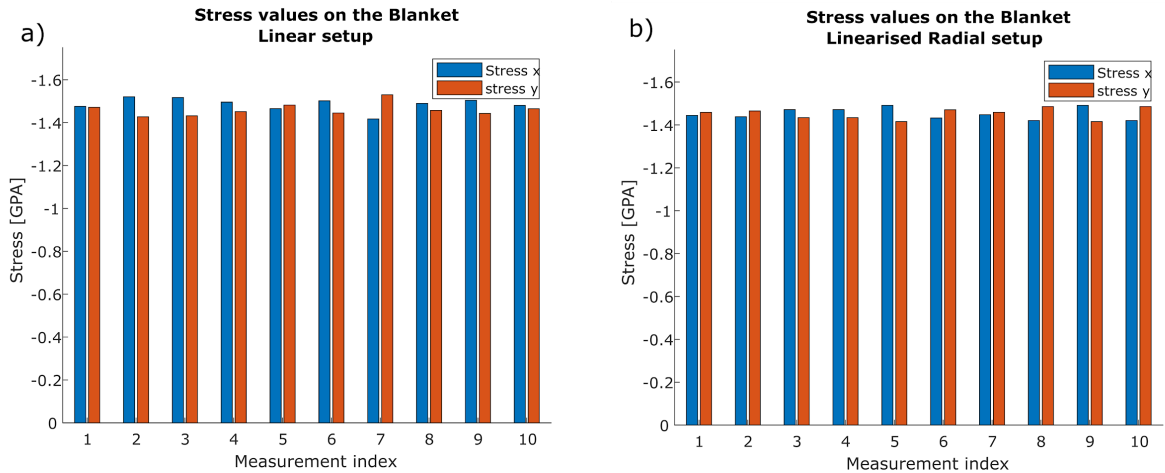

*Figure 11 Stress measured on the blanket structure on different locations near the center of the blanket structure using a) linearly polarised light setup and the b) linearised radially polarised light setup. Note the clear indication of biaxial nature of stress in these structures.*

Figure 11 illustrates the stress measured on the blanket structure using linearly polarised (FIgure 11a) and linearised radially polarised (Figure 11b). The average stress in the x and y directions are summarised in Table 5. Linearised radially polarised incoming light is used in the experimental radial setup by placing a linear polariser after the radial polariser.

*Table 5 Stress values measured used linearly polarised and linearised radially polarised incoming light setup on the Ge layer of the blanket structure. The error values on the stress measurements quoted here are the standard deviations obtained from multiple measurements.*

| Polarisation | Stress $\sigma_x$ [GPa] | Stress $\sigma_y$ [GPa] |
|---|---|---|
| Linear in x | $-1.47 \pm 0.05$ | $-1.48 \pm 0.05$ |
| Linearised-radial | $-1.45 \pm 0.03$ | $-1.45 \pm 0.03$ |

The error values on the stress measurements quoted here from both the linearly polarised and the linearised-radially polarised setups are the standard deviations obtained from multiple measurements as shown in Figure 11. These measurements, as expected, indicate biaxial in-plane compressive stress of equal magnitude in the blanket structure. The linearised-radially polarised setup clearly shows better precision (calculated as standard deviation in the stress values) in the stress calculation in comparison to the linearly polarized setup due to the increase in the longitudinal component of the incoming light which is expected to reduce the uncertainty in detecting the TO peak. This reduction in uncertainty also translates to better accuracy for the linearised-radially polarised setup as the measured stress values $\sigma_x$ and $\sigma_y$ are equal to each other on average as expected for a biaxially stressed Ge on top of a strain relaxed buffer.

To confirm these results, TEM strain measurements using the diffraction-based techniques like nano-beam diffraction[13,16,17,41] and Bessel diffraction [18] are performed, which are known to provide very good accuracy and precision for nano-scale strain measurements. The measurements are done on two perpendicular cross-section TEM lamellae prepared along the x and y directions. The strain is calculated with respect to a reference Si substrate in the xz and yz planes (Figure 12). The TEM diffraction experiments are performed using a Thermo Fisher Titan[3] aberration-corrected microscope operating at 300 kV. The convergence angle used for NBD was ~0.2 mrad and ~6 mrad for Bessel diffraction. The accuracy as measured for both NBD and Bessel on a bulk unstrained Ge sample (with known lattice parameters) using Si as reference was 2 x 10$^{-3}$. The precision measured as the standard deviation of strain over the reference Si area on the two perpendicular lamellae under the same illumination conditions is 7 x 10$^{-4}$ for NBD and 9 x 10$^{-4}$ for Bessel. The precision measured for Bessel is lower in comparison to our previously reported values [44] and the variations can be attributed to the thickness of the sample under investigation, the electron dose used [45] for acquisition and hence the overall SNR (signal to noise ratio) of the dataset. Hence, a determination of optimal electron dose level prior to the measurement to get good SNR aids to achieve the best possible precision depending on thickness of the sample of interest. However, Xianlin Qu and Qingsong Deng[46] have reported dominant knock-on beam damage in Si for higher electron dose rates around $1.8 - 2.6 \times 10^{20} e^- cm^{-2} s^{-1}$ and this can result in detrimental effects for strain measurement. The estimated dose for our experiment using a beam current of 35 pA and a total scanned area of approximately 450 x 100 nm² is 4.86 x $10^{17} \, e^- \, cm^{-2} \, s^{-1}$ and is well below the reported limits for beam damage.

Based on these stress values we can compare the overall strain as measured with Raman and TEM in Table 6. Note that this strain is computed in the Ge region with respect to bulk unstrained Ge (since the strain with respect to the bulk Ge material is of interest). The strain values reported from the Raman measurements are calculated from the measured stress values using the theory of elasticity and Hooke's law[32,47]. Analysing the normal strain values in the three perpendicular directions from TEM and Raman (Table 6) we find a close correspondence between the biaxial compressive normal strain in the x and y directions. However, the observed tensile strain in the z direction ($\varepsilon_{zz}$) is due to the Poisson effect and the $\varepsilon_{zz}$ measured with TEM is lower than the calculated value from the Raman data.

The difference could be explained by the strain relaxation inherent to the thin TEM lamella preparation (< 200 nm thickness). The relaxation is predominantly along the thickness direction versus the width of the lamella, as the latter is much larger (~1 to 2 µm). Obviously, this relaxation will influence the tensile strain in the z direction. This predominant relaxation through the thickness of the lamella is believed to lead only to tensile strain relaxation in the z direction and hence, it is not possible to measure it directly. Indeed since the TEM diffraction analysis is done on the projection through the thickness of the sample, it is not possible to decipher the strain parallel to the electron propagation direction with the current analysis. Hence, two perpendicular cross section lamellae are used (Figure 12) to evaluate normal strains along the three crystal coordinates x = [110], y = [− 110] and z = [001]. The exact quantification of the relaxation effect requires more complicated analysis using finite element method (FEM) simulations of the device structure[48] and will not be discussed in this paper.

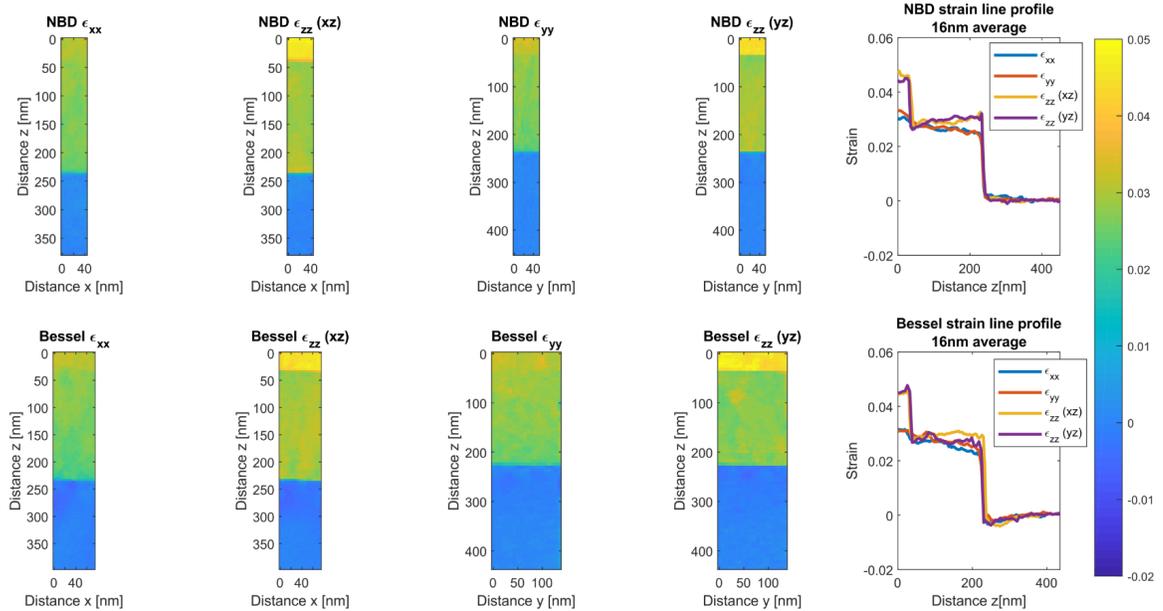

*Figure 12 Strain maps on the blanket structure from Nano-beam diffraction(NBD) and Bessel diffraction. The strain maps shown are from two perpendicular cross section lamellae in the xz and yz planes. The line profiles drawn are averaged horizontally over a distance of 16 nm.*

*Table 6 Strain values in the three perpendicular directions in strained Ge region with respect to bulk Ge as reference. The values obtained from the Raman are calculated from the measured stress values using Hooke's law and stress-strain relations in the elastic regime. The $\varepsilon_{zz}$ values from the TEM measurements are an average of the individual measurements in the xz and yz planes from the two perpendicular cross section lamellas. The uncertainty values are a measure of standard deviation from the observed strain values in the data.*

| Technique | $\varepsilon_{xx}$ | $\varepsilon_{yy}$ | $\varepsilon_{zz}$ |
|---|---|---|---|
| Bessel diffraction | $(-11 \pm 1) \times 10^{-3}$ | $(-10 \pm 1) \times 10^{-3}$ | $(3 \pm 1) \times 10^{-3}$ |
| Nano-beam diffraction (NBD) | $(-10.9 \pm 0.7) \times 10^{-3}$ | $(-9 \pm 0.9) \times 10^{-3}$ | $(3.3 \pm 0.6) \times 10^{-3}$ |
| Raman linearly polarised setup | $(-10.2 \pm 0.3) \times 10^{-3}$ | $(-10.1 \pm 0.3) \times 10^{-3}$ | $(7.5 \pm 0.2) \times 10^{-3}$ |
| Raman linearised-radially polarised setup | $(-10.2 \pm 0.2) \times 10^{-3}$ | $(-10.5 \pm 0.2) \times 10^{-3}$ | $(7.6 \pm 0.1) \times 10^{-3}$ |

We also compare the two variants of Raman spectroscopy utilized, namely using the oil immersion lens with linearly polarised incoming light and the linearised-radially polarised incoming light as applied for measurement on 16 nm-wide finFET structures (Figure 13). The 16 nm finFET structure is shown in Figure 14a. The schematic shows the building block of the finFETs array where the finFETs are 16nm wide and are distanced apart by a constant pitch of 200 nm. The building blocks are spatially repeated next to each other to form an array of 16nm finFET nanodevices. The device has a Si substrate and the STI (shallow trench isolation) first approach is followed in the device preparation[49]. The substrate is etched to create shallow trenches and a thick $Si_{0.3}Ge_{0.7}$ layer is selectively epitaxially grown (~110nm) inside the trenches[50]. A thin germanium channel(~30nm) is epitaxially grown on top of the $Si_{0.3}Ge_{0.7}$ with the lattice parameter of $Si_{0.3}Ge_{0.7}$. Hence, the Ge channel is expected to be compressively strained in the lateral direction since the lattice parameter of $Si_{0.3}Ge_{0.7}$ is smaller than that of Ge. The fins are separated by amorphous silicon oxide. The V-shape interface between the $Si_{0.3}Ge_{0.7}$ and the Si interface is used to trap the defects or dislocations from propagating into the Ge channel[51]. For the purpose of comparison of the two variants, the TO/LO ratio has been calculated by individually measuring the areas of the TO and LO peaks. The ratios are calculated over multiple measurements and the linearly polarised light yields a TO/LO ratio of $0.6 \pm 0.2$ while linearised-radially polarised light yields $1.2 \pm 0.3$. This again clearly illustrates that the linearised-radially polarised light provides a higher TO/LO ratio in comparison to linearly polarised light.

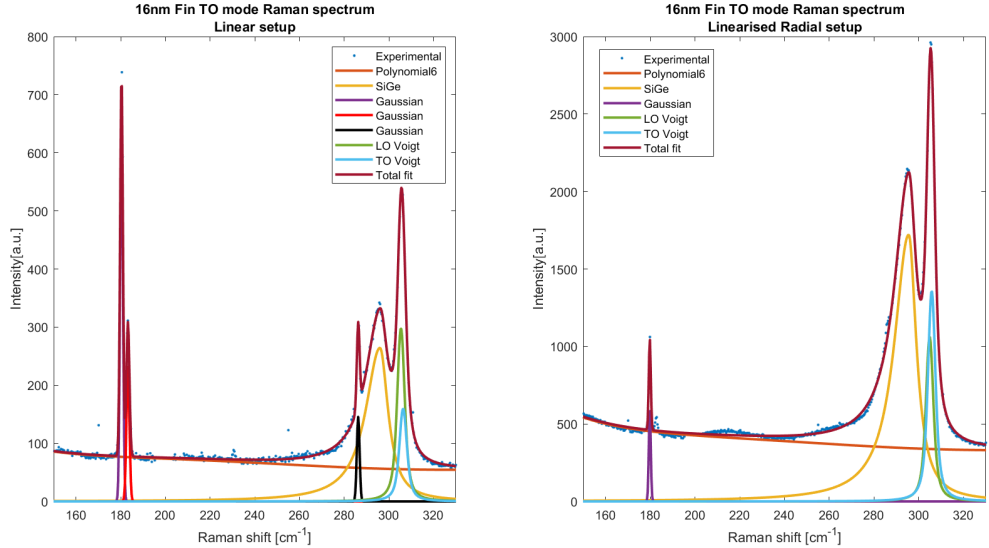

*Figure 13 Raman spectra obtained on a 16nm finFET using oil immersion lens from a linearly polarised and a linearised - radially polarised setups. The plasma lines are associated with the Rayleigh scattering of laser light and are positioned respectively at 180.2 cm$^{-1}$, 286.2 cm$^{-1}$ and so on and are fitted using the Gaussian profile[52].*

The stress measurements were also performed on the 16 nm finFET structure as shown in Figure 14a. Here, the stress $\sigma_y$ measured along the length of the finFET channel is higher in comparison to $\sigma_x$, reflecting the fact that these structures are designed to exhibit uniaxial stress along the channel direction. The average stress measured using linearly and linearised radially polarised light setup is summarised in Table 7.

*Table 7 Stress values measured used linearly polarised and linearised radially polarised incoming light setup on the 16 nm finFET Ge channel. The error values on the stress measurements quoted here are the standard deviations obtained from multiple measurements.*

| Polarisation | Stress $\sigma_x$ [GPa] | Stress $\sigma_y$ [GPa] |
|---|---|---|
| Linear in x | $-0.45 \pm 0.05$ | $-2.29 \pm 0.05$ |
| Linearised radial | $-0.48 \pm 0.02$ | $-2.26 \pm 0.02$ |

Note that while the absolute values for the stress measured using both techniques agree well, there is a clear difference in standard deviation between the two, already testifying to the advantage of using a radial polarizer in the setup. The measured standard deviations in the stress values are attributed to the different TO/LO ratios and in this regard, the linearised radially polarised setup clearly outperforms the linearly polarised setup. The same curve fitting procedure was used for the finFETs as in the case of blanket structures, but it was observed that the $Si_{0.3}Ge_{0.7}$ peak in TO-mode does not match with that in LO-mode for the finFET structures. This suggests that the $Si_{0.3}Ge_{0.7}$ is not completely relaxed as a strain relaxed buffer (SRB), and some residual stress is present in this layer too. TEM measurements are also taken with the two perpendicular cross section samples as shown in Figure 15. Here, we refer to the TEM sample cut along the xz plane of the finFET as the "cross-section lamella" and the sample cut along the yz plane as the "long-section lamella". The $\varepsilon_{yy}$ and $\varepsilon_{zz}$ line profile plots are not equal in the $Si_{0.3}Ge_{0.7}$ region as identified by both NBD and Bessel diffraction, which gives additional evidence for the assertion that the $Si_{0.3}Ge_{0.7}$ region is not completely stress and strain free. This would translate to a higher stress induction in the Ge channel which corresponds directly to the carrier mobility enhancement[53] and can also explain the higher $\varepsilon_{yy}$ strain values observed in the 16nm finFET in comparison to $\varepsilon_{yy}$ in the blanket structure.

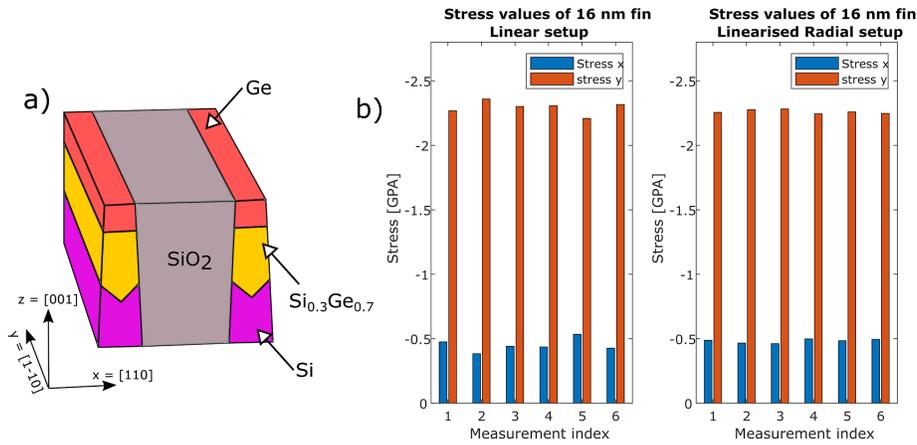

Figure 14 a) The schematic showing the building block of the finFETs array where the finFETs are 16nm wide and are distanced apart by a constant pitch of 200 nm. The building blocks are spatially repeated next to each other to form an array of 16nm finFET nanodevices. In Raman measurements, two to three building blocks are probed depending on the spatial resolution or the electric field distribution of light. b) Stress measured at different locations on the length of the structure using oil immersion lens and linearly polarised setup and the linearised radially polarised setup. Note the clear indication of uniaxial stress in these structures.

Finally, we report the measured strain values from TEM techniques and the calculated strain values from Raman for the Ge channel in the finFET in Table 8. We observe that the strain is almost relaxed in $\varepsilon_{xx}$ due to the reduced width of the finFET, a compressive strain $\varepsilon_{yy}$ along the length of the finFET and a tensile strain $\varepsilon_{zz}$ along the vertical or the growth direction [001] is observed due to the Poisson effect. Note that the $\varepsilon_{zz}$ values seen from TEM measurements are close to the calculated values from Raman. The apparently smaller relaxation effect for 16 nm finFETs in comparison to blankets can be understood by the fact that strain is already almost relaxed in $\varepsilon_{xx}$. For the TEM sample along the long section of finFET used to measure $\varepsilon_{zz}$ in our experiment (Figure 15), i.e. along the yz plane, the relaxation due to sample thinning would have a lower contribution in $x$ (through the thickness) since $\varepsilon_{xx}$ is almost relaxed prior to sample preparation.

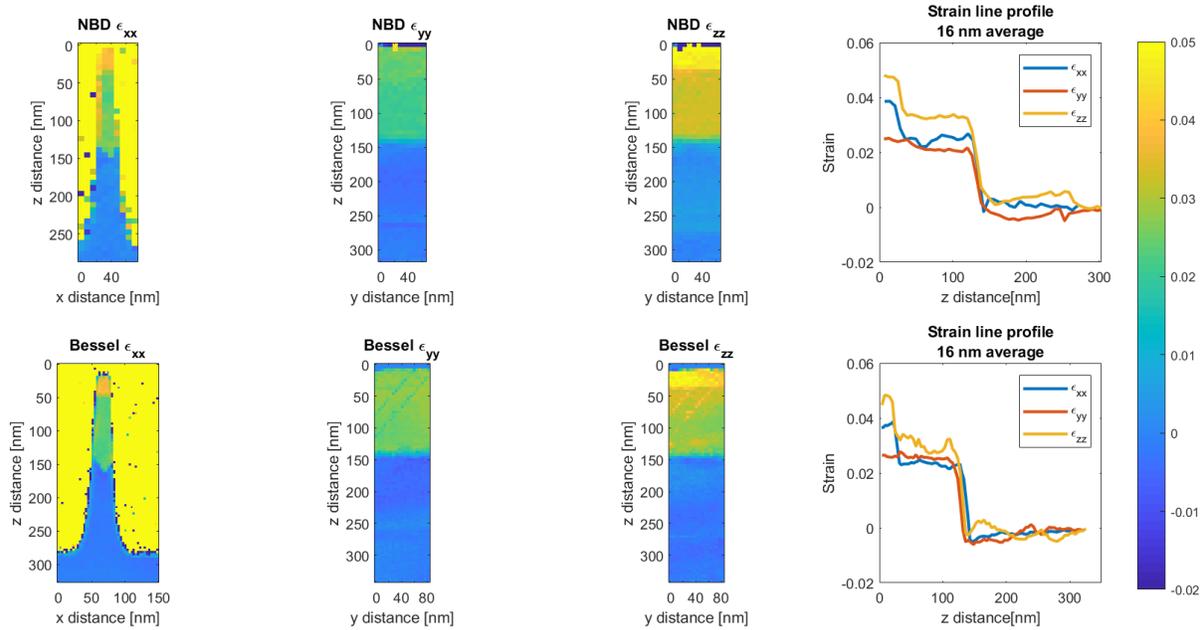

Figure 15 Strain measurement on a 16nm finFET. The $\varepsilon_{xx}$, $\varepsilon_{yy}$ and $\varepsilon_{zz}$ maps are drawn from two perpendicular cross section TEM lamellae. The line profiles are drawn vertically over the maps and are averaged horizontally over 16 nm.

The precision of the TEM techniques was also analysed for the measurements on the finFET (standard deviation on the reference Si region under identical acquisition conditions) and was found to be $9 \times 10^{-4}$ for both NBD and Bessel techniques. The $\varepsilon_{yy}$ and $\varepsilon_{zz}$ strain maps show a higher standard deviation in comparison to the precision analysed for the same measurement. This could be accounted for by the amorphous $SiO_2$ surrounding the FinFET. For the long section TEM sample in the yz plane (used to measure $\varepsilon_{yy}$ and $\varepsilon_{zz}$), the amorphous $SiO_2$ overlays with the crystal structure along the path of the travelling electrons (or through the thickness of the lamella) during measurement, while this situation is avoided for a cross section TEM lamella (xz plane). This results in an electron beam passing through both amorphous $SiO_2$ and the crystal structure for a long section TEM lamella resulting in more diffuse scattering from the amorphous region[54] and thus can explain the higher standard deviation in $\varepsilon_{yy}$ and $\varepsilon_{zz}$ strain maps. The strain values from the TEM measurements agree within the accuracy limits of the techniques (Table 6 and 8) which, as measured on a bulk Ge sample with known lattice parameters, is equal to $2 \times 10^{-3}$. The Raman measurements average over a much larger area (approximately 1 $\mu m^2$) than the TEM measurements and hence have lower standard deviations over their measurements. Hence, TEM measurements provide a more nanoscopic view into the strain distributions on the sample but require sample preparation and are more susceptible to strain relaxation.

Table 8 Strain values from the three perpendicular directions in the strained Ge region for the finfet structure. The values obtained from the Raman are calculated from the measured stress values using Hooke's law and stress-strain relations in the elastic regime. The uncertainty values are a measure of standard deviation from the observed strain values in the data.

| Technique | $\varepsilon_{xx}$ | $\varepsilon_{yy}$ | $\varepsilon_{zz}$ |
|---|---|---|---|
| Bessel diffraction | $(-4.1 \pm 0.8) \times 10^{-3}$ | $(-14 \pm 2) \times 10^{-3}$ | $(6 \pm 3) \times 10^{-3}$ |
| Nano-beam diffraction (NBD) | $(-3.5 \pm 0.9) \times 10^{-3}$ | $(-16 \pm 1) \times 10^{-3}$ | $(6 \pm 2) \times 10^{-3}$ |
| Raman linearly polarised setup | $(-2.8 \pm 0.4) \times 10^{-3}$ | $(-16.5 \pm 0.4) \times 10^{-3}$ | $(7.1 \pm 0.2) \times 10^{-3}$ |
| Raman linearised-radially polarised setup | $(-3.0 \pm 0.2) \times 10^{-3}$ | $(-16.4 \pm 0.2) \times 10^{-3}$ | $(7.1 \pm 0.1) \times 10^{-3}$ |

# Conclusion

We investigated Raman spectroscopy for the measurement of strain in semiconductor nano-devices, in particular when using different types of incoming light polarizations, namely radially and linearly polarised light and the combination of the two resulting in linearised radially polarised light. The linearised radially polarised light is shown to provide a higher TO/LO ratio which improves the sensitivity in determining the TO peak, which in turn leads to more precise stress measurements. Numerical simulations for a high NA oil immersion system show that the radial and linearised radially polarised light sources provide upto 1.3 to 1.5 times better averaged spatial resolution (in terms of averaged electric field distributions in the x,y,z direction in the focal plane) in comparison to the linearly polarised light. The Raman measurements are compared with TEM diffraction techniques and a good agreement in the strain values is seen within the accuracy limits of the TEM techniques except for the strain in the z-direction where the relaxation within the TEM lamella leads to lower strain values. The Raman technique provides a fast and efficient way for measuring stress-strain while requiring no additional sample preparation steps. As the strain measurements in Raman are averaged over a few 100 nm a more statistically relevant value is obtained whereas, the TEM techniques provide a more nanoscopic view of the strain distribution within a device with spatial resolutions down to 2 nm as illustrated with the analysis of a 16 nm finFET. The introduction of a linearised - radially polarised setup is straightforward as it only requires installing an s-waveplate and a linear polariser directly in the incoming laser path of a traditional micro - Raman setup operating with a linearly polarised laser source. Even though the alignment procedure for linearised-radially polarised light is slightly elaborate[55] and time consuming in comparison to a traditional linearly polarised Raman setup, we demonstrate through this work that linearised-radially polarised light provides an optimized Raman measurement both in terms of spatial resolution and in stress precision and accuracy.

# Acknowledgement:

The Qu-Ant-EM microscope and the direct electron detector used in the diffraction experiments was partly funded by the Hercules fund from the Flemish Government. This project has received funding from the GOA project "Solarpaint" of the University of Antwerp. J.V acknowledges funding from the European Union's Horizon 2020 research and innovation programme under grant agreement No 823717 – ESTEEM3.

# References:

1. Kim, Yong-Bin, "Challenges for Nanoscale MOSFETs and Emerging Nanoelectronics," Trans. Electr. Electron. Mater. **11**(3), 93–105 (2010).
2. V. K. Khanna, "Short-Channel Effects in MOSFETs," in *Integrated Nanoelectronics*, NanoScience and Technology (Springer India, 2016), pp. 73–93.
3. N. Horiguchi, B. Parvais, T. Chiarella, N. Collaert, A. Veloso, R. Rooyackers, P. Verheyen, L. Witters, A. Redolfi, A. De Keersgieter, S. Brus, G. Zschaetzsch, M. Ercken, E. Altamirano, S. Locorotondo, M. Demand, M. Jurczak, W. Vandervorst, T. Hoffmann, and S. Biesemans, "FinFETs and Their Futures," in *Semiconductor-On-Insulator Materials for Nanoelectronics Applications*, A. Nazarov, J.-P. Colinge, F. Balestra, J.-P. Raskin, F. Gamiz, and V. S. Lysenko, eds., Engineering Materials (Springer Berlin Heidelberg, 2011), pp. 141–153.
4. C.-F. Lee, R.-Y. He, K.-T. Chen, S.-Y. Cheng, and S.-T. Chang, "Strain engineering for electron mobility enhancement of strained Ge NMOSFET with SiGe alloy source/drain stressors," Microelectron. Eng. **138**, 12–16 (2015).
5. K. Rim, K. Chan, L. Shi, D. Boyd, J. Ott, N. Klymko, F. Cardone, L. Tai, S. Koester, M. Cobb, D. Canaperi, B. To, E. Duch, I. Babich, R. Carruthers, P. Saunders, G. Walker, Y. Zhang, M. Steen, and M. Ieong, "Fabrication and mobility characteristics of ultra-thin strained Si directly on insulator (SSDOI) MOSFETs," in *IEEE International Electron Devices Meeting 2003* (IEEE, 2003), p. 3.1.1-3.1.4.
6. Y.-C. Li, H.-M. Zhang, S. Liu, and H.-Y. Hu, "Strain and Dimension Effects on the Threshold Voltage of Nanoscale Fully Depleted Strained-SOI TFETs," Adv. Condens. Matter Phys. **2015**, 1–6 (2015).
7. J. Liu and J. M. Cowley, "High-resolution scanning transmission electron microscopy," Ultramicroscopy **52**(3–4), 335–346 (1993).
8. J. L. Rouvière, A. Mouti, and P. Stadelmann, "Measuring strain on HR-STEM images: application to threading dislocations in Al$_{0.8}$In$_{0.2}$N," J. Phys. Conf. Ser. **326**, 012022 (2011).
9. S. Kim, S. Lee, Y. Oshima, Y. Kondo, E. Okunishi, N. Endo, J. Jung, G. Byun, S. Lee, and K. Lee, "Scanning moiré fringe imaging for quantitative strain mapping in semiconductor

devices," Appl. Phys. Lett. **102**(16), 161604 (2013).
10. D. Su and Y. Zhu, "Scanning moiré fringe imaging by scanning transmission electron microscopy," Ultramicroscopy **110**(3), 229–233 (2010).
11. M. J. Hÿtch, E. Snoeck, and R. Kilaas, "Quantitative measurement of displacement and strain fields from HREM micrographs," Ultramicroscopy **74**(3), 131–146 (1998).
12. P. L. Galindo, S. Kret, A. M. Sanchez, J.-Y. Laval, A. Yáñez, J. Pizarro, E. Guerrero, T. Ben, and S. I. Molina, "The Peak Pairs algorithm for strain mapping from HRTEM images," Ultramicroscopy **107**(12), 1186–1193 (2007).
13. A. Béché, L. Clément, and J.-L. Rouvière, "Improved accuracy in nano beam electron diffraction," J. Phys. Conf. Ser. **209**, 012063 (2010).
14. L. Bruas, V. Boureau, A. P. Conlan, S. Martinie, J.-L. Rouviere, and D. Cooper, "Improved measurement of electric fields by nanobeam precession electron diffraction," J. Appl. Phys. **127**(20), 205703 (2020).
15. J.-L. Rouviere, A. Béché, Y. Martin, T. Denneulin, and D. Cooper, "Improved strain precision with high spatial resolution using nanobeam precession electron diffraction," Appl. Phys. Lett. **103**(24), 241913 (2013).
16. C. Mahr, K. Müller-Caspary, T. Grieb, M. Schowalter, T. Mehrtens, F. F. Krause, D. Zillmann, and A. Rosenauer, "Theoretical study of precision and accuracy of strain analysis by nano-beam electron diffraction," Ultramicroscopy **158**, 38–48 (2015).
17. P. Favia, M. Bargallo Gonzales, E. Simoen, P. Verheyen, D. Klenov, and H. Bender, "Nanobeam Diffraction: Technique Evaluation and Strain Measurement on Complementary Metal Oxide Semiconductor Devices," J. Electrochem. Soc. **158**(4), H438 (2011).
18. G. Guzzinati, W. Ghielens, C. Mahr, A. Béché, A. Rosenauer, T. Calders, and J. Verbeeck, "Electron Bessel beam diffraction for precise and accurate nanoscale strain mapping," Appl. Phys. Lett. **114**(24), 243501 (2019).
19. F. Houdellier, C. Roucau, L. Clément, J. L. Rouvière, and M. J. Casanove, "Quantitative analysis of HOLZ line splitting in CBED patterns of epitaxially strained layers," Ultramicroscopy **106**(10), 951–959 (2006).
20. D. Kosemura, S. Yamamoto, K. Takeuchi, K. Usuda, and A. Ogura, "Examination of phonon deformation potentials for accurate strain measurements in silicon–germanium alloys with the whole composition range by Raman spectroscopy," Jpn. J. Appl. Phys. **55**(2), 026602 (2016).
21. E. Anastassakis, A. Pinczuk, E. Burstein, F. H. Pollak, and M. Cardona, "Effect of static uniaxial stress on the Raman spectrum of silicon," Solid State Commun. **8**(2), 133–138 (1970).
22. A. Gawlik, J. Bogdanowicz, A. Schulze, J. Misiewicz, and W. Vandervorst, "Photonic properties of periodic arrays of nanoscale Si fins," in *Photonic and Phononic Properties of Engineered Nanostructures IX*, A. Adibi, S.-Y. Lin, and A. Scherer, eds. (SPIE, 2019), p. 76.
23. A. Gawlik, J. Bogdanowicz, T. Nuytten, A.-L. Charley, L. Teugels, J. Misiewicz, and W. Vandervorst, "Critical dimension metrology using Raman spectroscopy," Appl. Phys. Lett. **117**(4), 043102 (2020).
24. T. Nuytten, J. Bogdanowicz, T. Hantschel, A. Schulze, P. Favia, H. Bender, I. De Wolf, and W. Vandervorst, "Advanced Raman Spectroscopy Using Nanofocusing of Light: Advanced Raman Spectroscopy Using Nanofocusing…," Adv. Eng. Mater. **19**(8), 1600612 (2017).
25. Y. Saito, M. Kobayashi, D. Hiraga, K. Fujita, S. Kawano, N. I. Smith, Y. Inouye, and S. Kawata, "z -Polarization sensitive detection in micro-Raman spectroscopy by radially polarized incident light," J. Raman Spectrosc. **39**(11), 1643–1648 (2008).

26. A. Tarun, N. Hayazawa, H. Ishitobi, S. Kawata, M. Reiche, and O. Moutanabbir, "Mapping the "Forbidden" Transverse-Optical Phonon in Single Strained Silicon (100) Nanowire," Nano Lett. **11**(11), 4780–4788 (2011).
27. I. D. Wolf, "Micro-Raman spectroscopy to study local mechanical stress in silicon integrated circuits," Semicond. Sci. Technol. **11**(2), 139–154 (1996).
28. I. Loa, S. Gronemeyer, C. Thomsen, O. Ambacher, D. Schikora, and D. J. As, "Comparative determination of absolute Raman scattering efficiencies and application to GaN," J. Raman Spectrosc. **29**(4), 291–295 (1998).
29. R. Loudon, "The Raman effect in crystals," Adv. Phys. **13**(52), 423–482 (1964).
30. G. H. Loechelt, N. G. Cave, and J. Menéndez, "Polarized off-axis Raman spectroscopy: A technique for measuring stress tensors in semiconductors," J. Appl. Phys. **86**(11), 6164–6180 (1999).
31. I. De Wolf, "Relation between Raman frequency and triaxial stress in Si for surface and cross-sectional experiments in microelectronics components," J. Appl. Phys. **118**(5), 053101 (2015).
32. "Stress-strain equations," in *Handbook of Geophysical Exploration: Seismic Exploration* (Elsevier, 2003), **34**, pp. 61–68.
33. M. Erdelyi, J. Simon, E. A. Barnard, and C. F. Kaminski, "Analyzing Receptor Assemblies in the Cell Membrane Using Fluorescence Anisotropy Imaging with TIRF Microscopy," PLoS ONE **9**(6), e100526 (2014).
34. K. Bahlmann and S. W. Hell, "Electric field depolarization in high aperture focusing with emphasis on annular apertures," J. Microsc. **200**(1), 59–67 (2000).
35. K. S. Youngworth and T. G. Brown, "Focusing of high numerical aperture cylindrical-vector beams," Opt. Express **7**(2), 77 (2000).
36. R. H. Jordan and D. G. Hall, "Free-space azimuthal paraxial wave equation: the azimuthal Bessel–Gauss beam solution," Opt. Lett. **19**(7), 427 (1994).
37. K. S. Youngworth and T. G. Brown, "Inhomogenous polarization in scanning optical microscopy," in J.-A. Conchello, C. J. Cogswell, A. G. Tescher, and T. Wilson, eds. (2000), pp. 75–85.
38. R. J. Harrach, "Electric Field Components in a Focused Laser Beam," Appl. Opt. **12**(1), 133 (1973).
39. V. Poborchii, T. Tada, and T. Kanayama, "Enhancement of the Strained Si Forbidden Doublet Transverse Optical Phonon Raman Band for Quantitative Stress Measurement," Jpn. J. Appl. Phys. **51**, 078002 (2012).
40. J. M. Bennett, "POLARIZATION | Introduction," in *Encyclopedia of Modern Optics* (Elsevier, 2005), pp. 190–204.
41. T. S. Perova, J. Wasyluk, K. Lyutovich, E. Kasper, M. Oehme, K. Rode, and A. Waldron, "Composition and strain in thin Si1−xGex virtual substrates measured by micro-Raman spectroscopy and x-ray diffraction," J. Appl. Phys. **109**(3), 033502 (2011).
42. W. Liang, J. Li, and H. He, "Photo-Catalytic Degradation of Volatile Organic Compounds (VOCs) over Titanium Dioxide Thin Film," in *Advanced Aspects of Spectroscopy*, M. Akhyar Farrukh, ed. (InTech, 2012).
43. R. Ossikovski, Q. Nguyen, G. Picardi, and J. Schreiber, "Determining the stress tensor in strained semiconductor structures by using polarized micro-Raman spectroscopy in oblique backscattering configuration," J. Appl. Phys. **103**(9), 093525 (2008).
44. V. Prabhakara, D. Jannis, G. Guzzinati, A. Béché, H. Bender, and J. Verbeeck, "HAADF-STEM block-scanning strategy for local measurement of strain at the nanoscale," Ultramicroscopy **219**, 113099 (2020).

45. V. Prabhakara, D. Jannis, A. Béché, H. Bender, and J. Verbeeck, "Strain measurement in semiconductor FinFET devices using a novel moiré demodulation technique," Semicond. Sci. Technol. **35**(3), 034002 (2020).
46. X. Qu and Q. Deng, "Damage and recovery induced by a high energy e-beam in a silicon nanofilm," RSC Adv. **7**(59), 37032–37038 (2017).
47. M. A. Slawinski, ed., "Chapter 4 - Strain energy," in *Seismic Waves and Rays in Elastic Media*, Handbook of Geophysical Exploration: Seismic Exploration (Pergamon, 2003), **34**, pp. 69–84.
48. N. Cherkashin, T. Denneulin, and M. J. Hÿtch, "Electron microscopy by specimen design: application to strain measurements," Sci. Rep. **7**(1), 12394 (2017).
49. C.-P. Chang, "Shallow Trench Isolation," in *Encyclopedia of Materials: Science and Technology* (Elsevier, 2001), pp. 8437–8444.
50. F. Niu and B. W. Wessels, "Epitaxial growth and strain relaxation of BaTiO[sub 3] thin films on SrTiO[sub 3] buffered (001) Si by molecular beam epitaxy," J. Vac. Sci. Technol. B Microelectron. Nanometer Struct. **25**(3), 1053 (2007).
51. R. Loo, A. Y. Hikavyy, L. Witters, A. Schulze, H. Arimura, D. Cott, J. Mitard, C. Porret, H. Mertens, P. Ryan, J. Wall, K. Matney, M. Wormington, P. Favia, O. Richard, H. Bender, A. Thean, N. Horiguchi, D. Mocuta, and N. Collaert, "Processing Technologies for Advanced Ge Devices," ECS J. Solid State Sci. Technol. **6**(1), P14–P20 (2017).
52. M. Kakkala, "Vibrational spectroscopic studies of olivines, pyroxenes, and amphiboles at high temperatures and pressures," PhD Thesis, University of Hawaii at Manoa (1993).
53. "STRAIN EFFECTS ON HOLE MOBILITY OF SILICON AND GERMANIUM P-TYPE METAL-OXIDE-SEMICONDUCTOR FIELD-EFFECT-TRANSISTORS," (n.d.).
54. "Phases and Crystal Structures," in *Pergamon Materials Series* (Elsevier, 2007), **12**, pp. 1–86.
55. "Wophotonics," https://wophotonics.com/products/s-waveplate/.